# Efficient light coupling between an ultra-low loss lithium niobate waveguide and an adiabatically tapered single mode optical fiber

**NI YAO,**[1,7] **JUNXIA ZHOU,**[2,3,7] **RENHONG GAO,**[4,5] **JINTIAN LIN,**[4,8] **MIN WANG,**[2,3] **YA CHENG,**[2,3,4,5,6,*] **WEI FANG,**[1,†] **AND LIMIN TONG**[1]

[1] *Interdisciplinary Center for Quantum Information, State Key Laboratory of Modern Optical Instrumentation, College of Optical Science and Engineering, Zhejiang University, Hangzhou 310027, China*
[2] *State Key Laboratory of Precision Spectroscopy, East China Normal University, Shanghai 200062, China*
[3] *XXL—The Extreme Optoelectromechanics Laboratory, School of Physics and Electronics Science, East China Normal University, Shanghai 200241, China*
[4] *State Key Laboratory of High Field Laser Physics, Shanghai Institute of Optics and Fine Mechanics, Chinese Academy of Sciences, Shanghai 201800, China*
[5] *University of Chinese Academy of Sciences, Beijing 100049, China*
[6] *Collaborative Innovation Center of Extreme Optics, Shanxi University, Taiyuan, Shanxi 030006, China*
[7] *Ni Yao and Junxia Zhou contributed equally to this work.*
[8]*jintianlin@siom.ac.cn*
*\*ya.cheng@siom.ac.cn*
*†wfang08@zju.edu.cn*

**Abstract:** A lithium niobate on insulator ridge waveguide allows constructing high-density photonic integrated circuits thanks to its small bending radius offered by the high index contrast. Meanwhile, the significant mode-field mismatch between an optical fiber and the single-mode lithium niobate waveguide leads to low coupling efficiencies. Here, we demonstrate, both numerically and experimentally, that the problem can be solved with a tapered single mode fiber of an optimized mode field profile. Numerical simulation shows that the minimum coupling losses for the TE and TM mode are 0.32 dB and 0.86 dB, respectively. Experimentally, though without anti-reflection coating, the measured coupling losses for TE and TM mode are 1.32 dB and 1.88 dB, respectively. Our technique paves a way for a broad range of on-chip lithium niobate applications.

## 1. Introduction

Lithium niobate on insulator (LNOI) is an emerging platform which has shown promising potential for integrated photonics due to its numerous advantages of high refractive-index contrast, wide transparent window (0.35-5 μm), large nonlinear coefficient, and excellent electro-optic characteristics [1-5]. In recent years, photonic devices based on LNOI, including waveguides [6,7], microring resonators [1, 8], microdisk resonators [9-11], and photonic crystal cavities [12,13] have been developed. On-chip functionalities, such as second harmonic generation [13-18], electro-optic modulator [3,19-23], and optical frequency comb [24,25], have experienced rapid progress with the above-mentioned LNOI micro-/nano-devices. Importantly, many integrated on-chip lithium niobate (LN) devices developed recently already feature low propagation losses [6-8,10], which is critical for practical applications. However, some challenges are yet to be overcome.

One major obstacle of integrated LNOI photonics for the practical applications is the lack of efficient light coupling mechanism, especially between standard single-mode optical fibers and single-mode LN waveguides, which can greatly degrade the device performances [4,26]. Much effort has been devoted to improving the fiber-to-chip coupling efficiency [26-30].

Generally, grating couplers, considered to be an important coupling method, have the advantage of high alignment tolerance. However, large insertion losses hinder its further practical application [28,31]. The end-face coupling on a hybrid silicon nitride/LN platform suffers from large coupling loss (> 6 dB/facet) due to the serious mode mismatch between the optical fiber and microscale LN waveguide [27]. Recently, the on-chip LN bilayer inversely tapered mode size converter is demonstrated to improve the mode match, and the fiber-to-chip coupling losses are as lower as 1.7 dB/facet at 1550 nm wavelength through an optimized nanofabrication process [26]. Nevertheless, the nanofabrication technology is complex and expensive. Moreover, further reducing the coupling loss by the inversed taper appears challenging.

Here we demonstrate an efficient and cost-effective technique to solve the problem. The key of our technique is to adiabatically taper a standard single-mode fiber until the mode of the fiber sufficiently matches the mode of specially designed LN waveguide. Technically, the cross sectional profile of the fiber can be precisely controlled by our home-built fiber-pulling system [32,33]. Remarkably, our numerical simulation shows that the coupling loss can be as low as 0.32 dB and 0.86 dB for TE and TM mode via optimizing the taper diameter, respectively. By accurately coupling the fiber taper with the LN waveguide, coupling losses of 1.32 dB and 1.88 dB for TE and TM mode, respectively, have been measured in our experiment. The experimental coupling loss can be further reduced if an anti-reflection coating can be applied to the end facet of the LN waveguide. The results indicate that our coupling technique offers an effective means for engineering the mode field distribution.

## 2. The geometric parameters of the LNOI waveguide and fiber taper

The single-mode LNOI waveguides can be fabricated using a chemo-mechanical polish lithographic technique, which allows us to achieve a propagation loss of 0.042 dB/cm as reported in Refs. 6 and 7. As the technical details in the waveguide fabrication has been given therein, we only briefly describe the fabrication procedures below. The single-mode ridge waveguides were produced on a 700 nm-thick X-cut LN thin film which is first coated with a chromium (Cr) layer with a thickness of 600 nm by magnetron sputtering. The Cr layer was patterned into the stripe patterns using femtosecond laser direct ablation. The stripe waveguide pattern was transferred from the Cr mask to the underneath LN thin film by chemo-mechanical polishing (CMP). Afterwards, a chemical wet etching was carried out to remove the Cr mask, leaving behind the multi-mode ridge waveguides with a top width of ~1 μm. Single-mode waveguides with a suitable refractive index contrast were finally produced by coating a layer of $Ta_2O_5$ as the cladding layer of the waveguides [7]. The scanning electron microscopy (SEM) image of the fabricated LN ridge waveguide is presented in Fig. 1(a). The two end facets of the LNOI waveguides were smoothed using focused ion beam (FIB) milling. As shown by the SEM image in the inset of Fig. 1(a), the cross section of the waveguide exhibits a trapezoidal shape with a top width of 1.0 μm and a bottom width of 3.2 μm. As we have proved in our previous work, the waveguide can support single mode propagation at the communication wavelength of ~1.5 μm with a loss of ~0.04 dB/cm [7].

The high-quality fiber taper was produced by pulling a standard single-mode optical fiber (Corning SMF-28e) in a highly controlled manner. The home-made fiber-puller consists of a hydrogen flame torch, two motorized translation stages, and optical measurement components. Technical details of the fiber taper fabrication can be found in the Refs. 32 and33. The diameter of the fiber taper can be precisely controlled, and the transmittance is near unit (>98% for bi-conical taper). The high precision and reproducibility in the fabrication of fiber taper is critical for achieving a good mode-matching between the fiber taper and the LNOI waveguide. Once a fiber taper with waist diameter of 1.4 μm was fabricated, it was cut in half by FIB with a beam current of 260 pA at the position of desirable diameter which is determined by simulations. The cross section of the fabricated fiber taper is shown in Fig.

1(b), evidencing a fiber taper of a symmetric round shape with a diameter of ~1.4 μm. The fiber taper also shows a high surface smoothness with the FIB cutting.

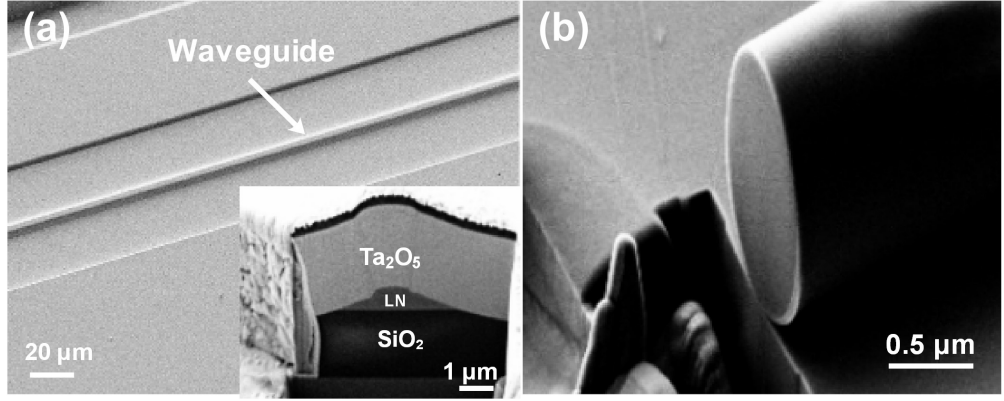

Fig. 1. (a) SEM image of the LNOI waveguides with a tilted view angle. Inset: SEM image of the cross section of waveguide. (b) SEM image of the tip of a fiber taper after FIB process.

## 3. Schematic of the integrated device and numerical simulations of the losses

Three-dimensional finite-difference time-domain (3D FDTD) simulations were performed to study optical mode profiles and coupling efficiency. Both TE and TM modes were calculated for fiber taper and LNOI waveguide, as shown in Fig. 2. Here the parameters of LNOI waveguide were set based on experimental structure. The refractive indices of top and bottom cladding layers were 2.05 and 1.444, respectively. And the refractive index of birefringent material lithium niobate were 2.211 ($n_o$) and 2.138 ($n_e$), respectively.

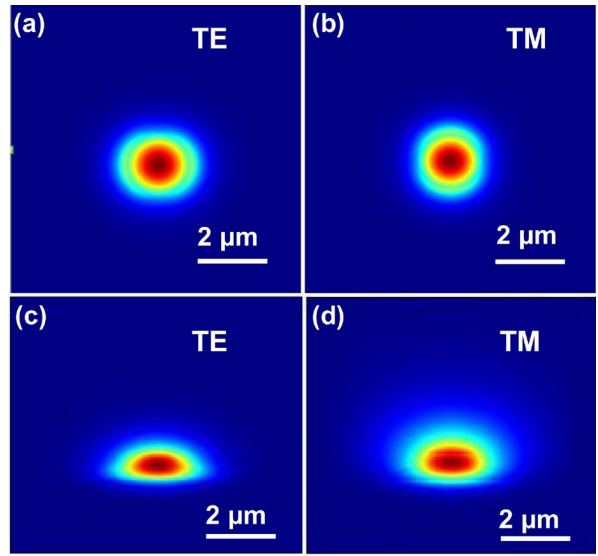

Fig. 2. Simulated mode field distributions in the fiber taper and waveguide, including the field distributions of (a) TE and (b) TM modes of the fiber taper with a diameter of 1.4 μm, as well as the field distributions of (c) TE and (d) TM modes of the LNOI waveguide in Fig. 1(a).

The coupling efficiency between the fiber taper and LNOI waveguide fiber mode field was estimated based on the structure schematically illustrated in Fig. 3(a). Specifically, an anti-reflection coating was applied on the input facet of waveguide, and the fiber taper with a diameter of $D$ direct contacted waveguide as butt-coupling. A 1550-nm laser with fundamental taper mode profile was launched into the fiber taper, and coupled to waveguide to excite the fundamental mode of the waveguide. The coupling efficiency was then calculated by comparing the power coupled out of the fundamental mode of the waveguide with that coupled into the fiber taper. This configuration allows us to determine the upper limit of the coupling efficiency between the fiber taper and waveguide, which provides the indication of the potential of our technique. TE and TM polarized modes were calculated separately, as shown in Fig. 3(b) and 3(c). The coupling loss and coupling efficiency as functions of fiber taper diameter for TE mode are shown in Fig. 3(b). Here each data point was get after careful adjusting the fiber taper position to reach optimal coupling. The minimum coupling loss was observed when the fiber taper diameter was 1.4 μm. For the TM mode, the coupling loss and coupling efficiency were calculated while the position of fiber taper was the same as that in Fig. 3(b) for each taper diameter. The coupling loss of TM mode drops monotonically from 1.04 dB to 0.76 dB when the fiber taper diameter increases from 1.0 μm to 2.0 μm, showing a different trend comparing to that of the TE mode. Here we mainly focused in TE mode, while the optimization could be carried out in the similar way for TM mode. Taking the fiber taper diameter as 1.4 μm, we obtained a coupling efficiency of 93% and 82% for TE and TM modes, respectively. Further improvements on the coupling efficiency are still possible by optimizing the waveguide parameters or shaping the cross section of the fiber taper into an elliptical shape.

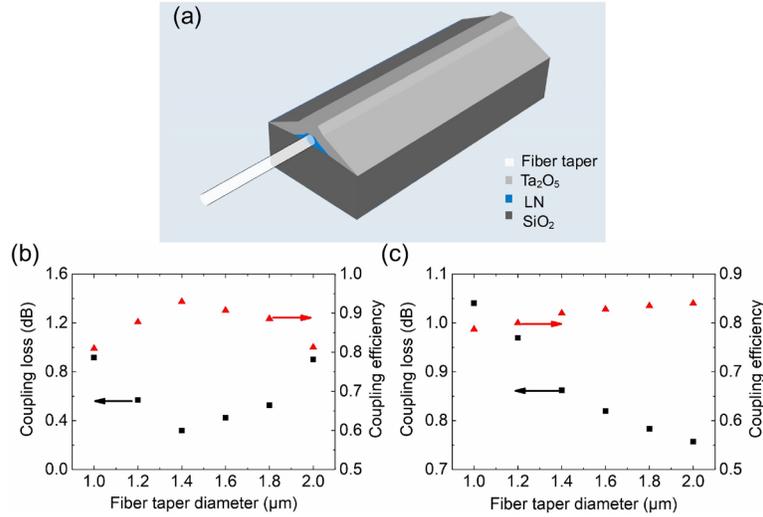

Fig. 3. (a) Schematic diagram of the fiber taper coupled LN waveguide. (b) Optimal coupling loss (black squires) and coupling efficiency (red triangles) of the TE mode as a function of fiber taper diameter when the mode overlapping between the fiber taper and waveguide is optimized. (c) Coupling loss (black squires) and coupling efficiency (red triangles) of the TM mode as a function of fiber taper diameter for the same coupling condition as in (b).

## 4. Experimental results and discussions

The schematic diagram of the optical measurement setup is shown in Fig. 4(a). A wavelength tunable laser (New Focus Inc., Model TLB 6728) with a linewidth of 200 kHz from 1520 to 1570 nm was used as the light source. The fabricated fiber taper was mounted on a high-resolution positioning stage with six independent travel degrees of freedom: X, Y, Z, Pitch, Yaw, and Roll. To eliminate the vibration of the fiber taper, a low refractive index $MgF_2$ plate was used to support the taper section of the fiber taper while introducing least mode leakage. The polarization of the input light was continuously tuned by an online polarization controller. The light transmitted from the waveguide was first collimated by an objective lens (Mitutoyo Inc., Model: M Plan Apo NIR HR) with NA of 0.35 and then focused onto either a power meter (Coherent Inc., Model: OP-2 IR) or an infrared CCD (HAMAMATSU Inc.,

InGaAs camera). This arrangement allowed us to monitor the output power as well as capture the output pattern from the waveguide. To achieve the maximum coupling efficiency, the relative position and orientation between the fiber taper and the input facet of the waveguide were carefully adjusted. Since the translation stage used in the experiment can be operated at a resolution of 1 nm in XYZ direction, the optimal positioning can be ensured by looking for the maximum output power at the end facet of the LNOI waveguide. The optical microscopy image of the fiber taper coupled with the waveguide is shown in Fig. 3(b), where the fiber taper is in direct contact with the LNOI waveguide. The output pattern of TE and TM modes at 1550 nm wavelength were captured by the infrared CCD, as shown in Fig. 4(c) and Fig. 4(d), respectively. The spatial intensity distributions clearly indicate that only the fundamental TE and TM modes were guided in the waveguide.

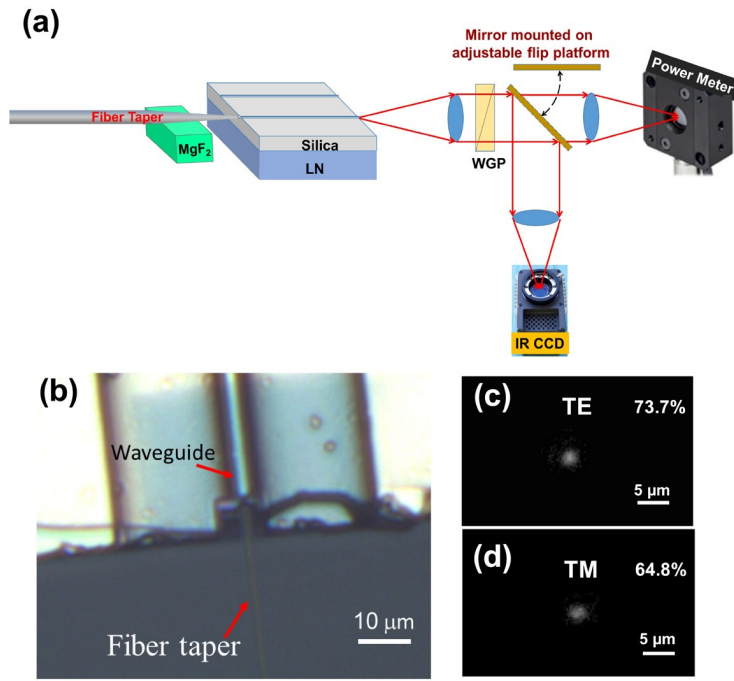

Fig. 4. (a) Optical measurement setup for the coupling between the fiber taper and waveguide. WGP, wire-grid polarizers. (b) The optical microscopy image of the fiber taper coupling with the waveguide. The output pattern of TE and TM modes were shown in (c) and (d), respectively.

Figure 5 plots the measured coupling loss as a function of input wavelength ranging from 1520 to 1570 nm when the diameter of fiber taper was fixed at 1.4 μm. Broadband response property is clearly shown with the coupling loss deviation less than 0.02 dB over the whole wavelength range, which is favorable for applications. The lowest coupling loss for TE mode was measured to be 1.32 dB at 1550 nm wavelength. Meanwhile, the corresponding coupling loss for TM mode measured at the same coupling condition was measured to be 1.88 dB. The experimental results agree well with the numerical simulation, considering the loss caused by the Fersnel reflection on the uncoated input and output facets of the LNOI waveguide in the experiment.

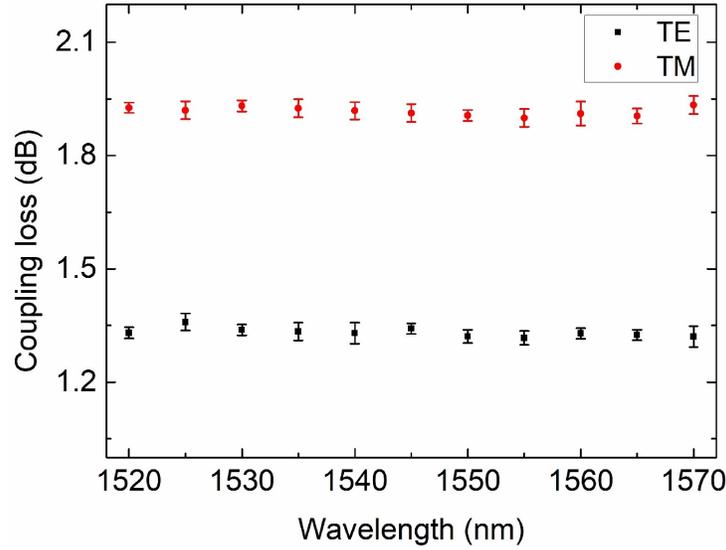

Fig. 5. The coupling efficiency of TE mode (black squares) and TM mode (red circles) for a fiber taper with a diameter of 1.4 μm. Error bars show the fluctuation in the measurements.

The high coupling efficiencies demonstrated in our experiment result from a few characteristic features of the LNOI waveguide and fiber taper. On the one hand, the mode field diameter of the waveguide is around ~2.5 μm. This is achieved by choosing a suitable index contrast between the cladding and core materials. Such a mode field size provides a relatively large tolerance of the misalignment of the fiber taper while still maintaining a reasonably small bend radius of the waveguide for high-density photonic integration applications. On the other hand, the fiber taper with a diameter of ~1.4 μm ensures both the high fabrication reliability and reproducibility without any great technical difficulty. Actually, there is still room for further promoting the coupling efficiency using our technique. First of all, the Fresnel reflections at the end facet of LNOI waveguide can be reduced by applying an optical coating, as proved by our simulation results. Second, the mode field distribution in the LNOI waveguide is now asymmetric which can be improved in the future by refining the fabrication parameters and the cladding material. Although it is possible to shape the fiber taper into an elliptical cross section using some nanofabrication technologies, we expect that it may not be efficient and cost effective for mass production. Last but not least, the technique demonstrated above is scalable by replacing the individual fiber taper with a fiber taper array, which will enable controllable coupling of light beams into multi-port PIC devices.

## 5. Conclusion

In conclusion, we have successfully achieved highly efficient coupling between optical fibers and LN waveguides. The numerical simulations indicate that the minimum coupling loss for both TE and TM are 0.32 dB and 0.86 dB, respectively, and experimentally, the coupling loss for TE and TM are 1.32 dB and 1.88 dB, respectively. It has been confirmed by experiment that the high coupling efficiency can be maintained for a wavelength range of ~50 nm at the telecommunication wavelength. As the coupling loss is one of the key parameters of photonic devices, our technique will benefit a broad range of LNOI-based PIC applications.

## 6. Funding, acknowledgments, and disclosures

### 6.1 Funding

National Basic Research Program of China (2019YFA0705000); the National Key R&D Program of China (2018YFB2200400); National Natural Science Foundation of China

(NSFC) (11734009, 11874375, 61590934, 11874154, 11674340, 61761136006, 11822410, 61635009); Key Research Program of Frontier Sciences, CAS (QYZDJ-SSW-SLH010); Key Project of the Shanghai Science and Technology Committee (18DZ1112700, 17JC1400400); the Strategic Priority Research Program of Chinese Academy of Sciences (XDB16030300); and the Fund (2019GZKF03006) supported by State Key Laboratory of Advanced Optical Communication Systems and Networks, SJTU, China.

*6.2 Acknowledgments*

We thank Dr. Wei Wang and Miss Liying Chen from Zhejiang University for preparing of FIB milling.

*6.3 Disclosures*

The authors declare that there are no conflicts of interest related to this article.